\documentclass[%
 prd,reprint,
 amsmath,amssymb,
 aps,
]{revtex4-2}

\usepackage{graphicx}
\usepackage{dcolumn}
\usepackage{bm}

\usepackage{xcolor}

\newcommand{\la}{\lambda} 

\newcommand{\ka}{\kappa}
\newcommand{\f}{\phi}

\newcommand{\al}{\alpha}
\newcommand{\bt}{\beta}
\newcommand{\ga}{\gamma}
\newcommand{\de}{\delta}
\newcommand{\si}{\sigma}

\newcommand{\ee}{\end{equation}}
\newcommand{\eea}{\end{eqnarray}}
\newcommand{\be}{\begin{equation}}
\newcommand{\bea}{\begin{eqnarray}}

\newcommand{\pa}{\partial}
\newcommand{\Om}{\Omega}
\newcommand{\vep}{\varepsilon}
\newcommand{\vrho}{\varrho}

\newcommand{\re}[1]{(\ref{#1})}
\newcommand{\R}{{\rm I \hspace{-0.52ex} R}}

\newcommand{\eins}{1\hspace{-0.56ex}{\rm I}}

\begin{document}


\title{$SO(4)$ gauged $O(5)$ Skyrmion on $\R^4$}


\author{Francisco Navarro-L\'erida}
\email[]{fnavarro@ucm.es}
\affiliation{Departamento de F\'isica Te\'orica and IPARCOS, Ciencias F\'isicas, Universidad Complutense de Madrid, E-28040 Madrid, Spain}

\author{D. H. Tchrakian}
\email[]{tigran@stp.dias.ie}
\affiliation{School of Theoretical Physics, Dublin Institute for Advanced Studies, Burlington Road, Dublin 4, Ireland}

\affiliation{Department of Computer Science, Maynooth University, Maynooth, Ireland}


\date{\today}

\begin{abstract}
We have studied an $SO(4)$ gauged $O(5)$ Skyrmion on $\R^4$ which can be seen as a static soliton
in $4+1$ dimensions. This is a sequel of the known $SO(D)$ gauged $O(D+1)$ Skyrmions on $\R^D$ in
$D=2$ and in $D=3$, like which they are localised to an absolute scale and are topologically stable,
their energies being bounded below by the winding number. In the absence of an analytic proof of existence
some such solutions are constructed numerically. Two families of solutions are found, of which
only one possesses a gauge decoupling limit. The curvatures of both of these solutions decay
as $r^{-3}$, a property they share with the Yang-Mills instantons on $\R^4$.
\end{abstract}


\maketitle

\section{Introduction and preliminaries \label{section1}}
Since the work of Schroers~\cite{Schroers:1995he} on the $SO(2)$ gauged $O(3)$ Skyrmions on $\R^2$,
there has been considerable
interest in carrying out the corresponding analysis on the $SO(3)$ gauged $O(4)$
Skyrmions~\footnote{
By Skyrmion here is meant the soliton of the $O(D+1)$ Skyrme scalar on $\R^D$
\begin{equation*}
  \f^A=(\f^a,\f^{D+1})\ ;\ A=a,D+1\ ;a=1,2,\dots,D \, ,
\end{equation*}
with $|\f^A|^2=|\f^a|^2+(\f^{D+1})^2=1$, \label{eq_footnote_1}
which in $D=3$ is parametrised as $U=\f^4\eins+\f^a\si_a$ in the original work~\cite{Skyrme:1961vr}
of Skyrme.} on $\R^3$~\cite{Arthur:1996np,nonlinearity,Cork:2018sem,Cork:2021ylu,Cork:2023pft}. It is our aim here
to extend this analysis
to the next step of constructing an $SO(4)$ gauged $O(5)$ Skyrmion on $\R^4$, namely to $D=4$ case,
which points out the extension to arbitrary $D$.

With the exception of the solutions in $D=2$~\cite{Schroers:1995he} where analytic proofs of existence for solutions
subject to no symmetries are given in~\cite{YYang,YYangSpringer} and in~\cite{HH}, solutions in
$D=3$~\cite{Arthur:1996np,nonlinearity} are studied only subject to
radial symmetry where the resulting ordinary differential equations (ODE's) are solved numerically. Here too we restrict our study to the radially
symmetric case and construct the solutions numerically. In the absence of a compelling physical reason to
consider less symmetric solutions, $e.g.,$ axial symmetry, and given the complexity of the system,
this is a reasonable approach.

The solutions constructed here have finite energy and are topologically stable, stabilised by
the Skyrme--Chern-Pontryagin (SCP) charges defined in~\cite{Tchrakian:2024bqy} for $D\le 5$. These
charge densities enable the construction of the models in $D=2,3$ and here in $D=4$, by exploiting
Bogomol'nyi~\cite{bog} like inequalities.

To be self-contained and to fix the notation, we give a brief description of the SCP topological charge used here.
In $D$ dimensions, the definitions of the winding number density of $\f^A$, and its gauge covariantised version $\vrho_G^{(D)}$ for gauge
group $SO(D)$, are
\bea
\vrho_0^{(D)}&=&\vep_{i_1i_2..i_D}\vep^{A_1A_2..A_D A_{D+1}}\nonumber\\
&&\times
\pa_{i_1}\phi^{A_1}\pa_{i_2}\phi^{A_2}\dots\pa_{i_D}\phi^{A_D}\f^{A_{D+1}} \,,\label{vr0}\\
\varrho_G^{(D)}&=&
\vep_{i_1i_2..i_D}\vep^{A_1A_2..A_D A_{D+1}}\nonumber\\
&&\times
D_{i_1}\phi^{A_1}D_{i_2}\phi^{A_2}\dots D_{i_D}\phi^{A_D}\f^{A_{D+1}}\label{vrG}\,,
\eea
where the $SO(D)$ covariant derivatives $D_i\phi^a $ and $D_i\phi^{D+1}$ are defined as
\bea
D_i\phi^a&=&\pa_i\phi^a+A_i^{ab}\phi^b\quad a=1,2,\dots,D\label{D}\,,\\
D_i\phi^{D+1}&=&\pa_i\f^{D+1}\,,  \label{D+1}
\eea
$A_i^{ab}$ being the $SO(D)$ gauge connection. Summation over repeated indices is assumed.

The $SO(D)$ gauge curvature is defined by $F^{ab}_{ij}\f^b=D_{[i}D_{j]}\f^a$, which explicitly reads 
\be
F_{ij}^{ab}=\pa_{[i}A_{j]}^{ab}+(A_{[i}A_{j]})^{ab}\, ,\label{curv}
\ee
where
\be
(A_{i}A_{j})^{ab}=A_{i}^{ac}A_j^{cb}  \quad a,b=1,2,\dots,D \, .
\ee

The method of construction of Skyrme--Chern-Pontryagin densities, which are
{\it both gauge-invariant and total-divergence}, is given in Ref.~\cite{Tchrakian:2024bqy}.

The prescription for constructing the SCP densities hinges on evaluating the difference
\[
\vrho^{(D)}_G-\vrho^{(D)}_0\,,
\]
and entails casting it in the form
\be
\label{diff}
\vrho^{(D)}_G-\vrho^{(D)}_0=\pa_i\Om^{(D)}_i[A,\f]-W^{(D)}[F,D\f]\,,
\ee
in which $W^{(D)}[F,D\f]$ is {\it gauge-invariant} and the vector valued density $\Om^{(D)}_i[A,\f]$ is {\it gauge-variant}.

From \re{diff}, one can define the density
\bea
\vrho&\stackrel{\rm def.}=&\vrho^{(D)}_G+W^{(D)}[F,D\f]\label{gi}\\
&\stackrel{\rm def.}=&\vrho^{(D)}_0+\pa_i\Om^{(D)}_i[A,\f] \,,\label{td}
\eea
which is both {\it gauge-invariant}, \re{gi}, and {\it total-divergence}, \re{td}, since $\vrho^{(D)}_0$ is essentially 
a total-divergence.

The definition \re{gi} can be exploited for constructing Bogomol'nyi like lower bounds, while the volume integral of \re{td} gives
the topological charge, which equals the {\it winding number}, as the surface integral of $\Om^{(D)}_i$
vanishes.

In Section \ref{section2} , the model under study is derived. In Section \ref{section3}, spherical symmetry is imposed and the resulting one dimensional system
is presented and the boundary values are fixed. Section \ref{section4} is devoted to the numerical construction of the solutions and a summary and
discussion is presented in Section \ref{section5}.

\section{The $SO(4)$ gauged $O(5)$ Skyrme model on $\R^4$ \label{section2}}
To emphasise the restriction to $\R^4$, we convert all $D$-dimensional Latin indices $i,j,\dots$, in the previous Section, to
$4$-dimensional Greek indeces $\mu,\nu,\dots$ here.

The quantities $W^{(D)}$ and $\Om_\si^{(D)}$ appearing in \re{gi}-\re{td} for $D=4$ are~\cite{Tchrakian:2024bqy}
\bea
W^{(4)}&=&\frac{3!}{4}\,\vep_{\mu\nu\rho\si}\vep^{fgab}\,\f^5\nonumber \\
&&\times
\bigg[
\frac{1}{6}(\f^5)^2\,F_{\mu\nu}^{fg}F_{\rho\si}^{ab}+F_{\mu\nu}^{fg}\f^{ab}_{\rho\si}\bigg]\,,\label{W4}\\
\Om_\si^{(4)}&=&\f^5\bigg(\pa_\mu\Om_{\mu\si}+\frac12\vep_{\mu\nu\rho\si}\vep^{fgab}
\bigg[F_{\mu\nu}^{fg}\f^aD_\rho\f^b \nonumber\\
&&  
+\frac12\left(1-\frac13(\f^5)^2\right)\Xi_{\mu\nu\rho}^{fgab}\bigg]\bigg)\,,\label{Oml4}
\eea
where $\f_{\mu\nu}^{ab}$ in \re{W4} denotes
\be
\label{fmn}
\f_{\mu\nu}^{ab}=D_{[\mu}\f^a D_{\nu]}\f^b\quad,\quad [\mu\nu]=-[\nu\mu] \,,
\ee
and in \re{Oml4}
\bea
\Om_{\mu\si}&=&\vep_{\mu\nu\rho\si}\vep^{fgab}A_\rho^{fg}\,\f^a
\left(\pa_\nu\f^b+\frac12A_\nu\f^b\right)\,,\label{tildeOm4}\\
\Xi_{\mu\nu\rho}^{fgab}&=&A_\rho^{ab}\left[\pa_\mu A_\nu^{fg}+\frac23(A_\mu A_\nu)^{fg}\right]\label{Xi4}\,.
\eea

$W^{(4)}$ defined by \re{W4} is the density in \re{gi} for $D=4$, which can be used to state
Bogomol'nyi like inequalities leading to the energy density functional that describes the soliton, namely \re{gi} for $D=4$
\be
\vrho^{(4)} = \vrho^{(4)}_G+W^{(4)}\,.\label{vr4}
\ee

However, \re{W4} is
not the most general such a density as indeed in all even dimensions the Chern-Pontryagin (CP) density, 
which is itself both {\it gauge invariant} and {\it total divergence}, can be added/subtracted
to both definitions of \re{gi} or \re{td} without changing these {\it properties} in both these. This has been exploited in the $D=2$
  case in \cite{Navarro-Lerida:2018siw} and \cite{Romao:2018egg}. In particular in Appendix {\bf A} of Ref.~\cite{Navarro-Lerida:2018siw},
it is explained why the inclusion of the CP density is necessary for supporting finite energy solutions.

In the present case, this is the $2$-nd CP density
\bea
&&-\frac14\,\vep_{\mu\nu\rho\si}\vep^{fgab}\,F_{\mu\nu}^{fg}F_{\rho\si}^{ab}=\nonumber \\
&& \vep_{\mu\nu\rho\si}\vep^{fgab}\,\pa_\si\left\{A_\rho^{fg}
\left[\pa_\mu A_\nu^{ab}+\frac23(A_\mu A_\nu)^{ab}\right]\right\}\,, \label{CP4}
\eea
the left hand side (with negative sign) being added to \re{gi} and the right hand side to \re{td}.

Adding the LHS of \re{CP4} to \re{W4}, and its RHS to \re{Oml4} we have the alternative definition of the SCP density in $D=4$
\bea
\tilde W^{(4)}&=&\frac{3!}{4}\,\vep_{\mu\nu\rho\si}\vep^{fgab}\nonumber \\
&&\times
\bigg\{
\frac{1}{6}[(\f^5)^3-1]\,F_{\mu\nu}^{fg}F_{\rho\si}^{ab}+\f^5\,F_{\mu\nu}^{fg}\f^{ab}_{\rho\si}\bigg\}\,,\label{W4tilde}\\
\tilde\Om^{(4)}_\si&=&\f^5\bigg(\pa_\mu\Om_{\mu\si}+\frac12\vep_{\mu\nu\rho\si}\vep^{fgab}\bigg[F_{\mu\nu}^{fg}\f^aD_\rho\f^b\nonumber \\
&&
+\frac12\left(3-\frac13(\f^5)^2\right)\Xi_{\mu\nu\rho}^{fgab}\bigg]\bigg)\,,\label{tildeOml4}
\eea
leading to the definition of the ``improved'' topological charge density
\be
\tilde\vrho^{(4)} = \vrho^{(4)}_G+\tilde W^{(4)}\,.\label{tildevr4}
\ee

Before proceeding to state the Bogomol'nyi like inequalities in the next Subsection, the choice of topological charge density
\re{vr4} or \re{tildevr4} must be made. Our choice here is the ``improved'' density \re{tildevr4} rather than \re{vr4}, because the
first term,
\be
[(\f^5)^3-1]\,F_{\mu\nu}^{fg}F_{\rho\si}^{ab} \,, \label{extra_piece}
\ee
in \re{W4tilde} leads to the definition of a (chiral) selfinteraction potential
in the same way as in the $D=2$ case~\cite{Schroers:1995he,Navarro-Lerida:2018siw,Romao:2018egg}. The difference is that in the $D=2$
case, it is necessary to incorporate the selfinteraction potential in the topological charge density for it to lead to a finite
energy lower bound, while in $D=4$ this is not necessary. We have nonetheless opted here for the ``improved'' density \re{tildevr4}
as the presence of the potential facilitates the numerical construction, which is the main means to prove the existence here.

\subsection{Lower bounds and choice of model \label{section2.1}}
To reproduce the first term in $\tilde W^{(4)}$ defined by \re{W4tilde}, appearing in \re{tildevr4}, consider the inequality
\be
\label{eineq1}
\left|(\tau_1\,\eta^{-2}\,F(4)-\tau\,\eta^2\,[(\f^5)^3-1])\right|^2\ge0\, ,\quad [\eta]=L^{-1} \, ,
\ee
in which $F(4)$ is defined as
\be
\label{F(4)}
F(4)\stackrel{\rm def.}=\vep_{\mu\nu\rho\si}\vep^{abcd}F_{\mu\nu}^{ab}F_{\rho\si}^{cd}\,.
\ee

Defining further the ``potential''
\be
\label{pot}
V[\f^5]\stackrel{\rm def.}=[(\f^5)^3-1]^2\,,
\ee
the inequality \re{eineq1} reads
\bea
&&\left(\tau_1^2\,\eta^{-4}\,F(4)^2+\tau^2\,\eta^4\,V[\f^5]\right) \ge\nonumber\\
&&2\tau\,\tau_1\,\vep_{\mu\nu\rho\si}\vep^{abcd}\,[(\f^5)^3-1]\,F_{\mu\nu}^{ab}F_{\rho\si}^{cd}\,, \label{eineq1x}
\eea
the RHS of which reproduces the first term in \re{tildevr4}.

To reproduce the second term in $\tilde W^{(4)}$, consider
the inequality

\be
\label{ineq20}
\left|\left(\tau_2\,\f^5\, F_{\mu\nu}^{fg}-\frac{1}{2^2}\,\tau_3\,\vep_{\mu\nu\rho\si}\vep^{fgab}\f_{\rho\si}^{ab}\right)\right|^2\ge 0 \,,
\ee
which leads to
\bea
&&\tau_2^2\,(\f^5)^2\,|F_{\mu\nu}^{fg}|^2+\tau_3^2\,|\f_{\mu\nu}^{fg}|^2\ge \nonumber \\
&&\frac12\,\tau_2\,\tau_3\,\vep_{\mu\nu\rho\si}\vep^{fgab}\,\f^5\,
F_{\mu\nu}^{fg}\f_{\rho\si}^{ab}\,. \label{ineq2}
\eea

Adding $\frac18$ times \re{eineq1x} to $3$ times \re{ineq2} we have the inequality
\bea
&&\frac18(\tau_1^2\,\eta^{-4}\,F(4)^2+\tau^2\,\eta^4\,V[\f^5])+
3\,\tau_2^2(\f^5)^2\,|F_{\mu\nu}^{fg}|^2 \nonumber \\
&&+3\,\tau_3^2\,|\f_{\mu\nu}^{fg}|^2\ge\tilde W^{(4)}\,,\label{lbtildeW}
\eea
provided that
\be
\label{constr12}
\tau\,\tau_1=1\quad{\rm and}\quad\tau_2\,\tau_3=1\,,
\ee
where $\tilde W^{(4)}$ is the density appearing in the definition \re{tildevr4} of the topological charge density.

To complete the RHS of the topological charge density \re{tildevr4}, we need to add $\vrho_G^{(4)}$.
This can be achieved by invoking the inequality~\footnote{Alternatively, one can invoke the inequality
\begin{equation*}
\left|\f_{\mu\nu}^{AB}-\tilde\f_{\mu\nu}^{AB}\right|^2\ge 0\,,
\end{equation*}
in which $\tilde\f_{\mu\nu}^{AB}$ is defined as, 
\begin{equation*}
\tilde\f_{\mu\nu}^{AB}=\frac{1}{2^2}\vep_{\mu\nu\rho\si}\vep^{ABCDE}\f_{\rho\si}^{CD}\f^E\,,
\end{equation*}
such that $\f_{\mu\nu}^{AB}\tilde\f_{\mu\nu}^{AB}=2\,\vrho_G^{(4)}$.
We have eschewed this option of introducing $\vrho_G^{(4)}$ since in that case the resulting energy density does not feature a quadratic Skyrme
kinetic term $|\f_\mu^A|^2$. Also in that case Derrick scaling is not satisfied in the gauge decoupling limit \label{footnote_2}}
\be
\label{ineq4}
\left|\left(\tau_4\,\ka^{-1}\f_\mu^A-\tau_5\,\ka\,\tilde\f_\mu^A\right)\right|^2\ge0 \,,
\ee
in which the constant $\ka$ has dimension of length, $[\ka]=L$, and the quantity $\tilde\f_\mu^A$ is defined as
\be
\tilde\f_\mu^A\stackrel{\rm def.}=\vep_{\mu\nu\rho\si}\vep^{ABCDE}\f_\nu^B\f_\rho^C\f_\si^D\f^E\label{tildefmuA}\,.
\ee

The inequality \re{ineq4} now leads to
\be
\label{ineq4x}
\frac12\left(\tau_4^2\,\ka^{-2}|\f_\mu^A|^2+\tau_5^2\,\ka^2|\tilde\f_{\mu}^{A}|^2\right)\ge\vrho_G^{(4)}\, ,
\ee
provided that
\be
\label{constr3}
\tau_4\,\tau_5=1\, .
\ee

Finally, adding \re{lbtildeW} to \re{ineq4x} we end up with the energy density
\bea
    {\cal\tilde E}&=&\frac18(\tau_1^2\eta^{-4}\,F(4)^2+\tau^2\eta^4\,V[\f^5])+3\tau_2^2\,(\f^5)^2|F_{\mu\nu}^{fg}|^2 \nonumber \\
&&+3 \tau_3^2\,|\f_{\mu\nu}^{fg}|^2
+\frac12\left(\tau_4^2\,\ka^{-2}|\f_\mu^A|^2+\tau_5^2\,\ka^2|\tilde\f_{\mu}^{A}|^2\right)\label{acttilde}\,,
\eea
which is bounded from below by the topological charge density \re{tildevr4}
\be
\tilde\vrho^{(4)}=\vrho^{(4)}_G+\tilde W^{(4)}\,. \label{tildevr4_2}
\ee

It may be worth remarking that saturating the Bogomol'nyi like lower bounds stated by inequalities \re{eineq1}, \re{ineq20} and \re{ineq4}
results in the trivial solution, since the ensuing equations are overdetermined.

\subsection{Scaling \label{section2.2} }
The energy density \re{acttilde} contains 8 parameters $\{\tau,\tau_1, \dots, \tau_5,\eta,\kappa\}$, subject to 3 constraints given by \re{constr12} and \re{constr3}. However, after rescaling properly, the effective number of free parameters may be reduced to only 2 parameters. In order to show that, we introduce the following dimensionless barred quantities, together with the dimensionless parameters $\alpha$ and $\beta$

\bea
x^\mu = \tau_5 \kappa {\bar x}^ \mu \, , & \eta = \beta^{1/4} \tau^{-1/2} (\tau_5 \kappa)^{-1} \, , \\ \alpha = \tau_3^2 \, , &
A^{f g}_\mu =  (\tau_5 \kappa)^{-1} {\bar A}^{f g}_\mu \, , \\ F^{f g}_{\mu \nu} =  (\tau_5 \kappa)^{-2} {\bar F}^{f g}_{\mu \nu} \, ,& F(4) = (\tau_5 \kappa)^{-4} {\bar F}(4) \, , \\
\phi^A = {\bar \phi}^A \, , & \phi^A_\mu =  (\tau_5 \kappa)^{-1} {\bar \phi}^A_\mu \, , \\ \phi^{AB}_{\mu\nu} =  (\tau_5 \kappa)^{-2} {\bar \phi}^{AB}_{\mu\nu} \, , 
&{\tilde \phi}^A_\mu =  (\tau_5 \kappa)^{-3} {\bar {\tilde \phi}}^A_\mu \, .
\label{rescaling}
\eea
When this rescaling is introduced in \re{acttilde} we get
\bea
    {\cal\tilde E}&=& (\tau_5 \kappa)^{-4} \left\{ \frac18(\beta^{-1}\,{\bar F}(4)^2+\beta\,V[{\bar\f}^5]) \right. \nonumber \\
&&  +3\alpha^{-1}\,({\bar \f}^5)^2|{\bar F}_{\mu\nu}^{fg}|^2+3\,\alpha\,|{\bar \f}_{\mu\nu}^{fg}|^2 \nonumber \\
&&  \left.      
+\frac12\left(|{\bar \f}_\mu^A|^2+|{\bar {\tilde \phi}}_{\mu}^{A}|^2\right) \right\}\label{acttilde_new} \,,
\eea
so choosing units such that $\tau_5 \kappa=1$ we obtain the energy density we will use for numerics 
\begin{widetext}
\be
{\cal\tilde E}=  \frac18(\beta^{-1}\,F(4)^2+\beta\,V[\f^5])+3\alpha^{-1}\,(\f^5)^2|F_{\mu\nu}^{fg}|^2+3\,\alpha\,| \f_{\mu\nu}^{fg}|^2
+\frac12\left(|\f_\mu^A|^2+|\tilde\f_{\mu}^{A}|^2\right)\label{acttilde_final}\,,
\ee
\end{widetext}
where we have suppressed all upper-bars for economy.

\begin{figure}[b]
 \includegraphics[height=.35\textwidth, angle =0 ]{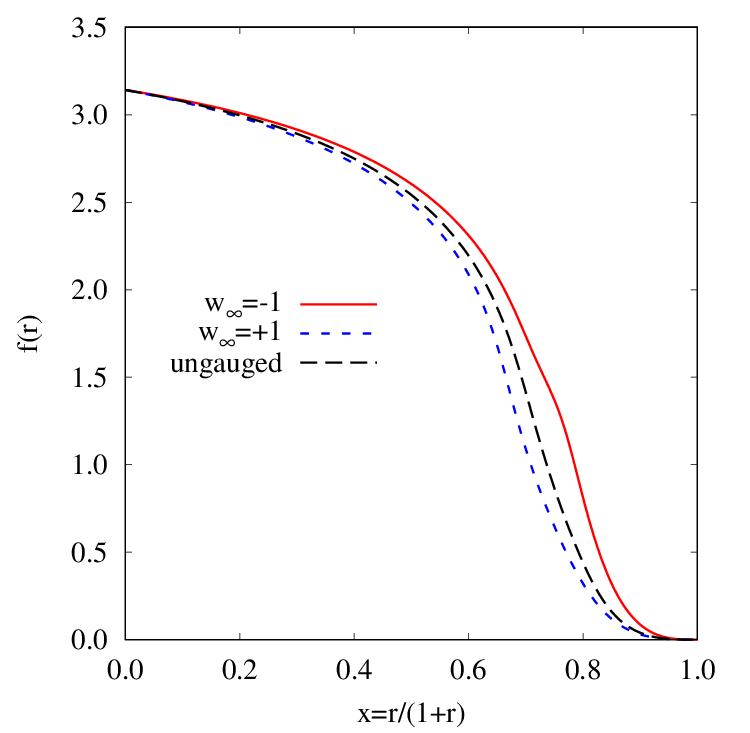} 
\caption{  
The profile of the function $f(r)$ for a typical solution is shown for $\alpha=\beta=1$, both for the gauged and the ungauged cases.
}
\label{fig1_new}
\end{figure}

\section{Imposition of radial symmetry \label{section3}}
Subject to spherical symmetry, the $SO(4)$ gauge connection on $\R^4$ is parametrised by a radial function $w(r)$ as by the Ansatz
\be
\label{redcon}
A_\mu^{ab}=-\left(\frac{1\pm w(r)}{r}\right)\de_\mu^{[a}\hat{x}^{b]}\ ;\quad  \hat{x}^{a}=\frac{x^a}{r}\quad a=1,2,3,4\,.
\ee

Subject to \re{redcon}, the $|F_{\mu\nu}^{ab}|^2$ term is 
\be
\label{F24}
|F_{\mu\nu}^{ab}|^2=\frac{12}{r^2}\left[(w')^2+\frac{1}{r^2}(1-w^2)\right]\,,
\ee
and the square of the $F(4)$ term \re{F(4)} is
\be
F(4)^2=(4\cdot 4!)^2\ \ \frac{(w')^2}{r^6}(1-w^2)^2\label{F4r2}\,.
\ee

Subject to this symmetry, the real valued (dimensionless) Skyrme scalar $\f^A=(\f^a,\f^5)$ is parametrised by the dimensionless
radial function $f(r)$ according to 
\bea
\f^a&=&\sin f(r)\,\hat{x^a} \, ,\label{Ska}\\
\f^5&=&\cos f(r)\,.\label{Sk5}
\eea

Denoting the $SO(4)$ covariant derivatives as
\be
\f_\mu^a\stackrel{\rm def.}=D_\mu\f^a\quad{\rm and}\quad\f_\mu^5\stackrel{\rm def.}=D_\mu\f^5\,, \label{D5}
\ee
subject to the Ansatz \re{redcon} and \re{Ska}-\re{Sk5}, \re{D5} are
\bea
\f_\mu^a&=&\left[\left(f'\cos f\pm\frac{w\sin f}{r}\right)\hat{x}_\mu\hat{x}^a\mp\frac{w\sin f}{r}\,\de_i^a\right] \,, \label{redcovh}\\
\f_\mu^5&=&-(f'\sin f)\,\hat{x}_\mu \,, \label{redcovg}
\eea
resulting in the quadratic Skyrme kinetic term
\be
\label{higgsquad24}
|\f_\mu^A|^2=(f')^2+\frac{3}{r^2}w^2\sin^2f\,.
\ee

\begin{figure}[b]
 \includegraphics[height=.35\textwidth, angle =0 ]{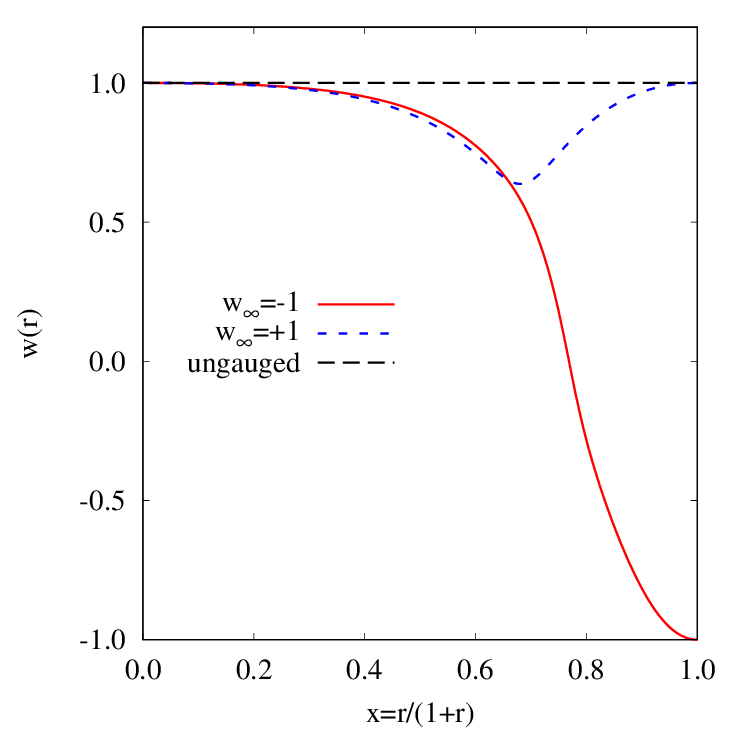} 
\caption{  
The profile of the function $w(r)$ for a typical solution is shown for $\alpha=\beta=1$, both for the gauged and the ungauged cases.
}
\label{fig2_new}
\end{figure}

The quartic~\footnote{This is not the term $|\f_{\mu\nu}^{AB}|^2$ which is invariant against the global $SO(5)$ rotations of the $O(5)$ model,
which is given by the simpler expression
\begin{equation*}
|\f_{\mu\nu}^{AB}|^2=2\cdot 3!\left(\frac{w\sin f}{r}\right)^2\left[(f')^2+\left(\frac{w\,\sin f}{r}\right)^2\right]\, , 
\end{equation*}
instead
}
Skyrme kinetic term $|\f_{\mu}^{ab}|^2$ in the energy \re{acttilde} is readily calculated to be
\be
\label{quarticz}
|\f_{\mu\nu}^{ab}|^2=2\cdot 3!\left(\frac{w\sin f}{r}\right)^2\left[(f'\cos f)^2+\left(\frac{w\,\sin f}{r}\right)^2\right]\,.
\ee

The sextic term in the energy density \re{acttilde} is $|\tilde\f_\mu^A|^2$, with $\tilde\f_\mu^A$ defined by \re{tildefmuA},
which yields
\bea
|\tilde\f^A|^2=|\f_{\mu\nu\la}^{ABC}|^2&=&(3!)^2\,\left(\frac{w\sin f}{r}\right)^4 \nonumber \\
&&\times \left[3(f')^2+\left(\frac{w\,\sin f}{r}\right)^2\right]\,. \label{sextic}
\eea

Substituting \re{redcon}, \re{Ska}, and \re{Sk5} into the ``energy'' density functional \re{acttilde_final}
and multiplying the latter with $r^3$, coming from the volume element, we have the one-dimensional ``energy'' density. The resulting one
dimensional ODE's are not displayed here, since they are quite cumbersome.

\subsection{Boundary conditions and expansions \label{section3.1}}

Substituting the spherical Ansatz \re{redcon}, \re{Ska}, and \re{Sk5} into the field equations derived variationally from \re{acttilde_final} subject to the constraint $|\phi^A|^2=1$, we obtain a system of two second-order equations for $f(r)$ and $w(r)$. Solutions to that system that have finite energy must satisfy
\bea
f(0)=\pi \, , & w(0)=1 \, ,  \label{BCs_origin} \\
\lim_{r\to\infty} f(r)=0 \, , & \displaystyle \lim_{r\to\infty} w(r)\equiv w_\infty=\pm 1 \, . \label{BCs_infty}
\eea

Note that the presence of the $|F_{\mu\nu}^{ab}|^2$ term prevents the existence of monopole-type solutions with
$ \displaystyle \lim_{r\to\infty} w(r)=0$, since that would lead to infinite-energy configurations, as seen clearly from \re{F24}.

Under \re{BCs_origin} and \re{BCs_infty}, the expansion of the solutions at the origin reads
\be
f(r) = \pi+ f_1 r + O(r^3) \, , \ w(r) = 1 +w_2r^2 + O(r^4) \, ,   \label{exp_origin}
\ee
while their asymptotic expansion results to be
\be
f(r) = \frac{{\hat f}_3}{r^3} + O\left(\frac{1}{r^5}\right) \, , \ w(r) = \pm 1 + \frac{{\hat w}_2}{r^2} +  O\left(\frac{1}{r^4}\right)  \, ,  \label{exp_infty}
\ee
where $f_1,w_2,{\hat f}_3,{\hat w}_2$ are constants determined numerically.

\section{Results \label{section4}}

We have solved the field equations derived from \re{acttilde_final}, subject to the spherical Ansatz, by using the software package COLSYS which employs a collocation method for boundary-value ordinary differential equations and a damped Newton method of quasi-linearization \cite{COLSYS}. The only free paramaters in this models are $\alpha$ and $\beta$, which we have varied in order to study the structure of the space of solutions.

The general form of the solutions that satisfy the expansions \re{exp_origin} and \re{exp_infty} is shown in Figs.~\re{fig1_new}-\re{fig2_new}, where the profiles of the functions $f(r)$ and $w(r)$ are displayed, respectively, for $\alpha=\beta=1$ for $w_\infty=\pm 1$ and the ungauged case, resulting from setting $w(r)=1$ in the field equations.

From the energy density  ${\cal\tilde E}$ \re{acttilde_final}, we can compute the energy of the solutions $E$ both for the gauged cases ($w_\infty=\pm 1$) and the ungauged one ($w(r)=1$) as
\be
E = \frac{1}{(4\pi)^2}\int {\cal\tilde E} d^4 x =  \frac{1}{32}\int r^3 {\cal\tilde E} dr \, ,\label{energy}
\ee
where we have extracted the angular contribution and also included a $1/32$ factor to normalize the topological charge $q$. The last equality in \re{energy} is valid for the spherical Ansatz \re{redcon}, \re{Ska}, and \re{Sk5}. By construction, \re{energy} has a lower bound given by the (absolute value of) the integral of $1/(4\pi)^2$ times \re{tildevr4_2}, which evaluated over the gauged ($w_\infty=\pm 1$) and the ungauged ($w(r)=1$) solutions takes the value
\be
|q| \equiv \left| \frac{1}{(4\pi)^2}\int \tilde\vrho^{(4)}  d^4 x    \right| = 1 \, .
\ee

In Fig.~\re{fig3_new} we represent the energy density ${\cal\tilde E}(r)$ as a function of the radial coordinate. Note that it presents its maximal value at the origin, for all the types of solutions considered here.

The analysis of the dependence of the energy $E$ on $\alpha$ and $\beta$ is carried out in the subsequent figures. In Figs.~\re{fig4_new}-\re{fig5_new} we fix one of the parameters and move the other one, for the gauged ($w_\infty=\pm 1$) and the ungauged ($w(r)=1$) solutions, presenting the results in logarithmic scale. The lower bound is also displayed. From these plots one can see that the energy of gauged $w_\infty=-1$ solutions is always larger than the corresponding one of the gauged $w_\infty=+1$ solutions. When compared to the energy of the ungauged solutions, whereas the gauged $w_\infty=+1$ solutions have always lower energy than the ungauged ones, the gauged $w_\infty=-1$ solutions have larger energy than the ungauged ones for small values of $\alpha$ and/or $\beta$, but their energy gets lower than that of the ungauged solutions beyond a certain value of the parameters. 

\begin{figure}[ht!]
\begin{center}
 \includegraphics[height=.35\textwidth, angle =0 ]{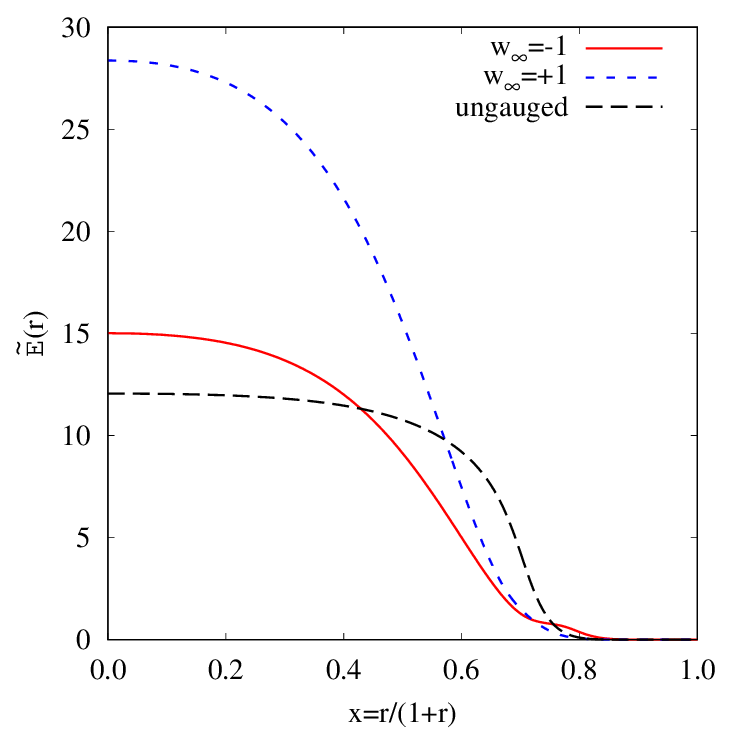}
\end{center}
\caption{  
Energy density ${\cal\tilde E}(r)$ for solutions with $\alpha=\beta=1$, both for the gauged and the ungauged cases.
}
\label{fig3_new}
\end{figure}

\begin{figure}[b]
 \includegraphics[height=.35\textwidth, angle =0 ]{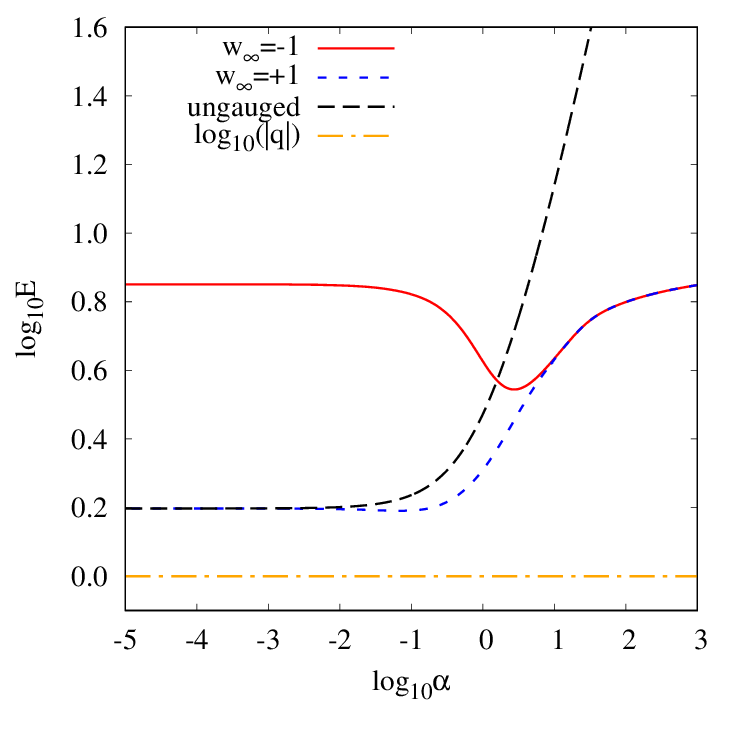} 
\caption{  
Energy $E$ versus the parameter $\alpha$ for fixed $\beta=0.1$, in logarithmic scale.}
\label{fig4_new}
\end{figure}

\begin{figure}[b]
 \includegraphics[height=.35\textwidth, angle =0 ]{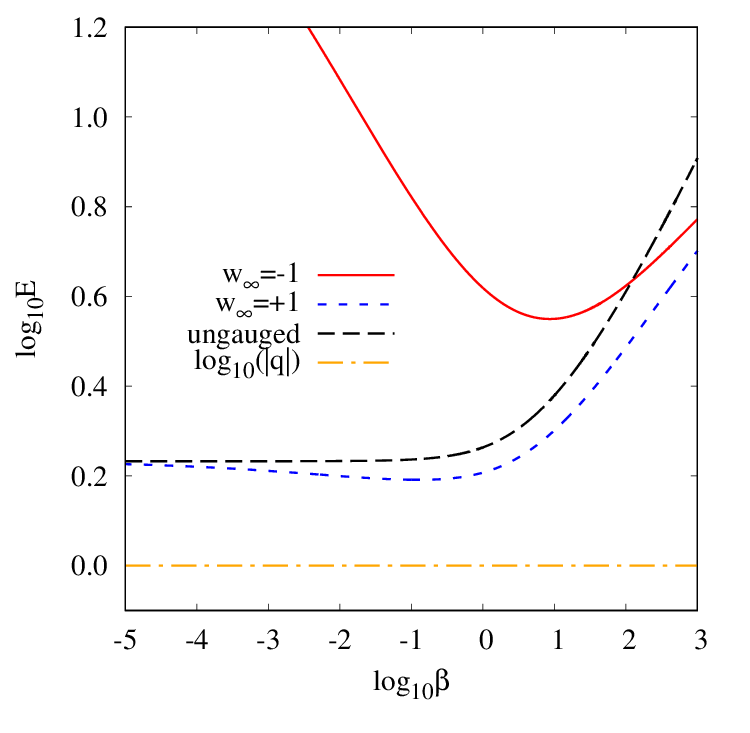}
\caption{  
Energy $E$ versus the parameter $\beta$ for fixed $\alpha=0.1$, in logarithmic scale.}
\label{fig5_new}
\end{figure}

\begin{figure}[b]
 \includegraphics[height=.35\textwidth, angle =0 ]{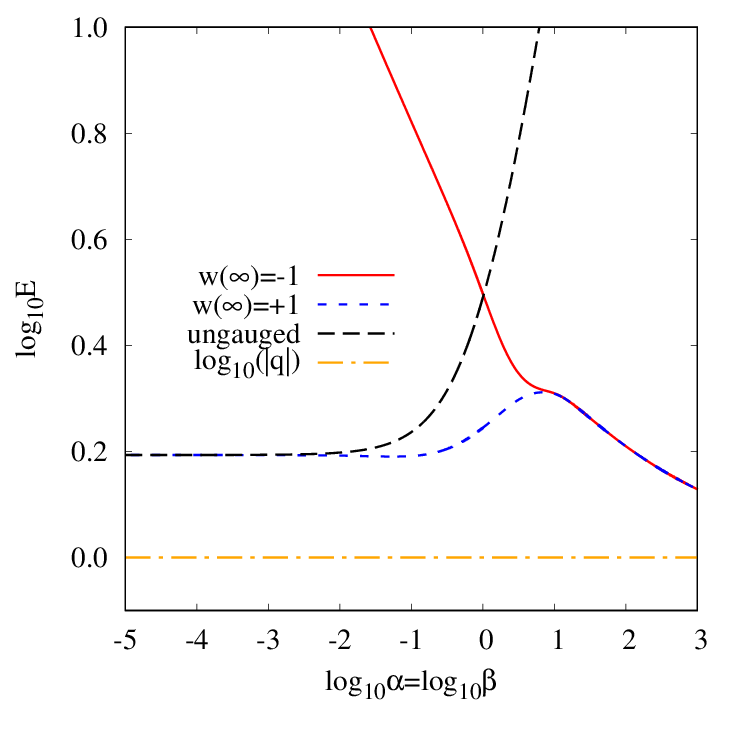} 
\caption{  
  Energy $E$ versus the parameter $\alpha=\beta$.}
\label{fig6_new}
\end{figure}

\begin{figure}[ht!]
 \includegraphics[height=.33\textwidth, angle =0 ]{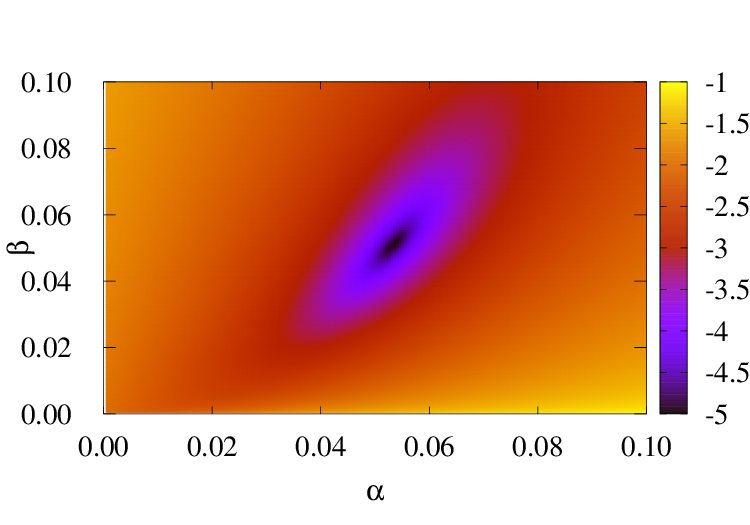}
\caption{  
  Detail of the local minimum of the energy of gauged $w_\infty=+1$ solutions, where $\log_{10}(E-1.55)$ is shown on the $(\alpha,\beta)$ plane.}
\label{fig7_new}
\end{figure}

Although the analysis of the full $(\alpha,\beta)$-parameter space is computationally too costly, we can further support our previous statements by considering the $\alpha=\beta$ set of solutions. Their energy is represented as a function of $\alpha=\beta$ in Fig.~\re{fig6_new}. In this plot it is clearly seen one feature also present in Figs.~\re{fig4_new}-\re{fig5_new}, namely, that in the large $\alpha$ and/or $\beta$ limit, the energies of gauged $w_\infty=-1$ and $w_\infty=+1$ solutions tend to coincide. In fact, for $\alpha=\beta$ they approach the lower bound more and more in the large $\alpha=\beta$ limit. As a final comment, we will indicate that the gauged $w_\infty=+1$ solutions show a local (not global) minimum of the energy, which is absent for the gauged $w_\infty=-1$ solutions and the ungauged ones. It is approximately located at $(\alpha,\beta)=(0.055,0.055)$, as graphically shown in Fig.~\re{fig7_new}.

\begin{figure}[b]
 \includegraphics[height=.35\textwidth, angle =0 ]{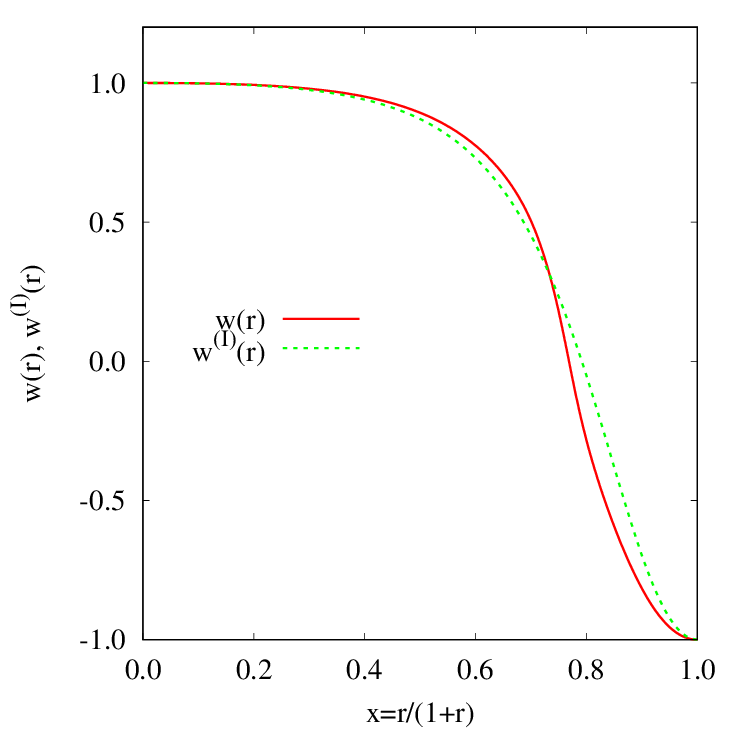}
\caption{  
Functions $w(r)$ and $w^{(I)}(r)$ for the solution with $\alpha=\beta=1, w_\infty=-1$ and the BPST solution for $\lambda=0.26334$, respectively. 
}
\label{fig8_new}
\end{figure}

\begin{figure}[b]
 \includegraphics[height=.35\textwidth, angle =0 ]{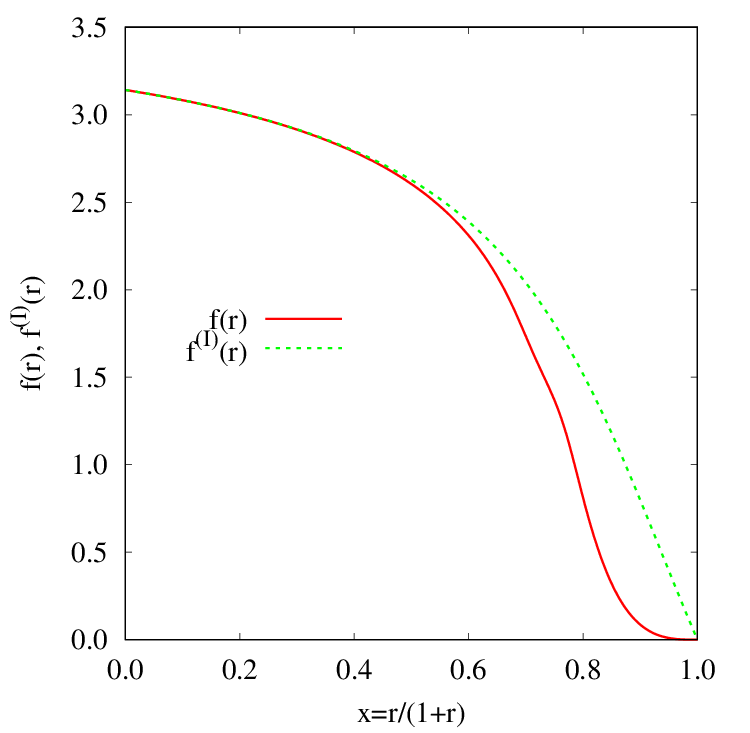} 
\caption{  
Functions $f(r)$ and $f^{(I)}(r)$ for the solution with $\alpha=\beta=1, w_\infty=-1$ and the BPST solution for $\lambda=0.26334$, respectively. 
}
\label{fig9_new}
\end{figure}

\section{Summary and discussion \label{section5}}

We have studied an $SO(4)$ gauged $O(5)$ Skyrme model on $\R^4$ in some detail, including quantitative aspects numerically. This is a sequel
to the seminal work~\cite{Schroers:1995he} of the $SO(2)$ gauged $O(3)$ Skyrme model on $\R^2$, and the corresponding
work~\cite{Arthur:1996np,nonlinearity,Cork:2018sem,Cork:2021ylu} of the $SO(3)$ gauged $O(4)$ Skyrme model on $\R^3$.
As such it points out the path to the construction of such models for all $D$, namely, $SO(D)$ gauged $O(D+1)$ Skyrme models supporting 
solitons on $\R^D$.
The most striking differences between the $D=2$ case~\cite{Schroers:1995he} and the $D=3$ case~\cite{Arthur:1996np,nonlinearity} as well as here
in the $D=4$ case is that in $D=2$ solutions to the first-order equations saturating the topological lower bound exist, while in $D=3$ and $4$
the corresponding ``Bogomol'nyi equations'' are overdetermined and only solutions to the full second-oder equations exist. Moreover, while
analytic proofs for the existence in $D=2$ are given in~\cite{YYang,YYangSpringer,HH}, in $D=3,4$ the existence of solutions to the
second-order ODE's subject to radial symmetry are given numerically.

Having claimed that the $D=4$ example presented here points the way to the case of arbitrary $D$, it is perhaps appropriate to point out that
the number of free parameters characterising the models increases with increasing $D$.
The $D=3$ case studied in~\cite{nonlinearity} features only one
parameter, say $\ka$, while here in $D=4$ the parameter space is described by two numbers $(\al,\bt)$, as formulated in
Section \ref{section2.2}.

There is yet another aspect of the prolification of choice of models with increasing $D$. Here in $D=4$ we
have chosen the energy density functional \re{acttilde} and have eschewed the option alluded to in footnote [2]
since 
in that case the model would not feature a quadratic Skyrme kinetic term.
It is clear such proliferation of choices will increase with increasing $D$.

A notable property of solutions in $D=3$ studied in~\cite{nonlinearity} and those studied here in $D=4$ is that there occur
two branches of solutions in each case. In $D=3$, the mass/energy $E$ versus $\ka$ plot $(E,\ka)$~\cite{nonlinearity} features a swallow's tail~\cite{Brihaye:1998vr}, while here in $D=4$
no such (swallow's tail) structures are observed in the $(E,\al)$ and $(E,\bt)$ profiles. In both $D=3$ and $D=4$ cases the two branches of solutions are
characterised by the asymptotic value of the function $\displaystyle w_{\infty}\stackrel{\rm def.}=\lim_{r\to\infty}w(r)$
and in both cases {\it only} one of these branches is connected to the gauge decoupling limit of the Skyrmion.

In $D=3$ each of these two branches terminate at a finite value of $\ka$, say $\ka_1$ and $\ka_2$, which
coexist in a certain range and are connected by a third branch that forms the swallow's tail. Only one
of these branches connects with the gauge decoupled Skyrmion and the whole $(E,\ka)$ profile lies below
the corresponding profile for the unguaged Skyrmion.

In $D=4$ by contrast the two branches, both parametrised by $(\al,\bt)$, are present in the whole range, and
do not stop anywhere in $\al$ and $\bt$. Moreover, only one of these is connected to the gauge
decoupled Skyrmion, and, as expected, its energy is always smaller than that of the ungauged Skyrmion.
The other branch, the one which does not have a gauge decoupling limit, can have energy larger
than the ungauged Skyrmion in some parts of parameter space.

It would be interesting to learn what is the source of this difference in the energy profiles in $D=3$
and $D=4$. A marked difference between even and odd dimensions is the presence of a Pontryagin
density, which enters the definition of the topological charge and results in the appearance of the
``pion mass'' type potential via the Bogomol'nyi like lower bounds. This is the case in both
$D=2$~\cite{Schroers:1995he} and here in $D=4$, in the first case there being a single branch and here
the two branches covering the whole parameter space. In $D=3$ by contrast, in the absence of a Pontryagin
density, no ``pion mass'' type potential appears naturally~\footnote{Should such a potential be added {\it by hand}
the situation may change but this would go beyond the minimal models implied by the Bogomol'nyi like inequalities.}.
This question can be settled by studying the $SO(5)$ gauged $O(6)$ Skyrmion on $\R^5$, which may be a subject
of a future investigation.

Finally, it is perhaps in order to point out a technical interpretation of the solutions constructed here
as instantons, by re-assigning the ``energy'' $E$ of the Skyrmion in $4+1$ dimensions as the ``action'' $S$ on $\R^4$.
Like the BPST instanton~\cite{Belavin:1975fg}, the two types of solutions here (characterised by the asymptotic
values $w_{\infty}=\pm1$ of the function $w(r)$) have a function $w(r)$ that decays as $r^{-2}$ and their actions are bounded below by
the topological charge, which is the winding number in this case. Unlike the BPST instantons, which satisfy
first-order equations saturating the lower bound, the solutions here do not saturate the lower bound but
in contrast to the former~\cite{Belavin:1975fg}, which are localised to an arbitrary scale, these are
localised to an absolute scale.

The difference between the BPST instanton~\cite{Belavin:1975fg} and
the solutions here is that the former are described by (anti-)self-dual $SU_{\pm}(2)$ connections, while
the solutions here are described by an $SO(4)$ connection, although between them there is a curious similarity.

The curvature $F_{\mu\nu}^{(\pm)}$ of the $SU_{\pm}(2)$ connection
\be
A_\mu^{(\pm)}=A_\mu^{ab}\,\si_{ab}^{(\pm)}\ ,\quad\si_{ab}^{(\pm)}=-\frac14\left(\frac{1\pm\ga_5}{2}\right)[\ga_\mu,\ga_\nu] \, , \label{SU2_curvature}
\ee
satisfying the first-order (anti-)self-duality equation
\be
F_{\mu\nu}^{(\pm)}=\pm\frac12\,\vep_{\mu\nu\rho\si}\,F_{\rho\si}^{(\pm)} \, , \label{SU2_selfduality}
\ee
yields, subject to spherical symmetry, the familiar charge-$1$ BPST (anti-)instanton.

The solutions to \re{SU2_selfduality} happen to coincide with those to the double--self-duality equations
\be
\label{adsd}
F_{\mu\nu}^{ab}=\pm\frac{1}{(2!)^2}\,\vep_{\mu\nu\rho\si}\vep^{abcd}F_{\rho\si}^{cd}\, ,
\ee
which for the radially symmetric $SO(4)$ connection \re{redcon}, yields the same function $w(r)$ that parametrise the BPST (anti-)instantons.

The function $w^{(I)}(r)$ solving the double--self-duality equation \re{adsd} coincides with the BPST function
\be
w^{(BPST)}(r)= \frac{1-\la^2r^2}{1+\la^2r^2}\, , \label{w_BPST}
\ee
is also a solution to the full second-order equations of the $SO(4)$ gauge field on $\R^4$.

But the function $w(r)$ describing
the second-order equations of the model \re{acttilde} differs from $w^{(I)}(r)$ since
the first-order double--self-duality equations saturating the inequalities \re{eineq1}, \re{ineq20} and \re{ineq4},
are overdetermined.

We have plotted the function $w^{(I)}(r)$ in Fig.~\ref{fig8_new} and the chiral
function~\footnote{The chiral function $f^{(I)}(r)$ here parametrises the $O(5)$
Skyrme scalar $\f^{\bar a}=(\f^a,\f^5)$ with
\begin{equation*}
\f^a=\hat x^a\,\sin f^{(I)}\ {\rm and}\ \ \f^5=\cos f^{(I)}\,, \label{eq_footnote_5_1}
\end{equation*}
satisfying the double-selfduality equation
\begin{equation*}
  \f_{\mu\nu}^{\bar a\bar b}=\pm\frac{1}{2^2}\vep_{\mu\nu\rho\si}\vep^{\bar a\bar b\bar c\bar d\bar e}\f_{\rho\si}^{\bar c\bar d}\f^{\bar e}\, ,
\end{equation*}  
where $\f_{\mu\nu}^{\bar a\bar b}=\pa_{[\mu}\f^{\bar a}\pa_{\nu]}\f^{\bar b}$
 }
\be
f^{(I)}(r)=\pi - \arccos w^{(I)}(r) \, , \label{f_I}
\ee
in Fig.~\ref{fig9_new}, to contrast them with the profiles of $w(r)$ and $f(r)$ that solve the second-order
equations of the model \re{acttilde}, for the $\alpha=\beta=1$, $w_\infty=-1$ case. For the BPST solution, $\lambda$ has been chosen such that $df/dr(0)=df^{(I)}/dr(0)$ (i.e., $\lambda=
0.26334$).

Although the main behaviour of the two types of solutions is similar, the main discrepancy occurs at infinity, where the presence of a potential (\ref{pot}) forces the function $f(r)$ in our solutions to decay as $1/r^3$ while for the BPST solution $f^{(I)}(r)$ decays as $1/r$.

Unfortunately the solitons with {\it instanton boundary values} presented here cannot be employed to construct a
{\it dilute gas} of instantons as in Section {\bf 3} of \cite{Polyakov:1976fu}. This is because the soliton employed in~\cite{Polyakov:1976fu}
is a {\it magnetic monopole} which describes a symmetry breaking field configuration, and  which allows gauging away of the order-parameter
(Higgs field), while the order-parameter (Skyrme scalar) here cannot be gauged away as that model does not describe a
{\it symmetry breaking} theory.

To enable the construction of a dilute gas of instantons on $\R^4$ a model of the $SO(4)$ gauged Higgs scalar $\f^a\,,\ a=1,2,3,4$,
must be employed. But in the spherically symmetric case the lower bounds analogous to \re{ineq4} and \re{ineq4x} result in {\it monopole
asymptotic behaviour}, which subject to the requirement of finite action excludes the presence of an $F^2$ curvature-square
term in the action density leaving only an $F^4$ term, which decays too fast and leads to the vanishing of the dilute gas action.
This obstacle can be overcome by seeking monopole--antimonopole solutions subject to bi-azimuthal symmetry (as $e.g.,$ in~\cite{Radu:2006gg})
that can describe finite action solutions with {\it instanton boundary conditions}, allowing the presence of an $F^2$ term
that on $\R^4$ supports a dilute gas. This is a challenge to which we plan to return.

\section*{Acknowledgements}
Special thanks to Eugen Radu for constant discussions during the preparation of the manuscript. We thank Yves Brihaye and Derek Harland for
valuable discussions and comments on the manuscript, and Yisong Yang for earlier discussions and for bringing
Refs.~\cite{YYang,YYangSpringer,HH} to our attention. We acknowledge helpful discussions with Ernesto Canfora, Bjarke Gudnason, and Parameswaran Nair.
One of us (D.H.Tch.) thanks Kieran Arthur for his early work on this subject.
F.N.-L. gratefully acknowledges support  from MICINN under project PID2021-125617NB-I00 ``QuasiMode".

\bibliography{paper_PRD_resub}

\end{document}